\documentclass[preprint,onecolumn,nofootinbib]{revtex4-1}
\pdfoutput=1
\usepackage[colorlinks=true,linkcolor=blue,urlcolor=blue,filecolor=black,citecolor=red,pdfstartview=FitV,pdftitle={},pdfsubject={},pdfkeywords={},pdfpagemode=None,bookmarksopen=true]{hyperref}
\usepackage{graphicx}
\usepackage{amsmath}
\usepackage{amsfonts}
\usepackage{amssymb,ulem}
\usepackage{color,xcolor}%
\usepackage{CJK}
\usepackage{subfigure}
\usepackage{dcolumn}
\newcommand{\f}{\begin{equation}}
\newcommand{\ff}{\end{equation}}
\newcommand{\fa}{\begin{eqnarray}}

\usepackage{booktabs}
\newcommand{\ffa}{\end{eqnarray}}

\begin{document}

\title{Universal geometric framework for black hole phase transitions: from multivaluedness to classification}
\author{Shi-Hao Zhang$^{1}$}
\author{Zi-Yuan Li$^{2}$}
\author{Jing-Fei Zhang$^{1}$}
\author{Xin Zhang$^{1,3,4}$}
\thanks{zhangxin@neu.edu.cn (corresponding author)}
\affiliation{
    $^1$ Liaoning Key Laboratory of Cosmology and Astrophysics, College of Sciences, Northeastern University, Shenyang 110819, China \\
    $^2$ School of Physics, Nankai University, Tianjin 300071, China \\
    $^3$ MOE Key Laboratory of Data Analytics and Optimization for Smart Industry, Northeastern University, Shenyang 110819, China\\
    $^4$ National Frontiers Science Center for Industrial Intelligence and Systems Optimization, Northeastern University, Shenyang 110819, China
}
\begin{abstract}
Recent studies have revealed synchronized multivalued behavior in thermodynamic, dynamical, and geometric quantities during the black hole first-order phase transition, which enables a diagnosis from different perspectives, yet its fundamental origin has remained poorly understood. By constructing a unified geometric framework integrating real analysis and covering space theory, we reveal the universal mathematical mechanism behind this phenomenon. We prove that this multivaluedness originates from two nondegenerate critical points in the temperature function $T(r_+)$, where $r_+$ is the horizon radius, which fold the parameter space into a three-sheeted covering structure. As a direct application, we propose that for the small/large black hole first-order phase transition in the canonical ensemble, a black hole undergoes such a transition if and only if its $T(r_+)$ curve has two extrema. Accordingly, we establish a classification scheme, denoted $A1$, $A2$, and $B$ for black holes. This scheme offers a complementary perspective to classifications based on global topological invariants. Our work provides a theoretical foundation for diagnosing phase transitions via multivaluedness and establishes a unified geometric perspective on black hole thermodynamics, chaotic dynamics, and spacetime structure during first-order phase transitions.

\end{abstract}

\maketitle
\tableofcontents

\section{Introduction}

As a gravitational system with thermodynamic properties, black holes exhibit rich phase structures. In particular, phase transitions in anti-de Sitter (AdS) spacetime offer profound insights into the nature of quantum gravity. Significant examples include the Hawking—Page phase transition between thermal AdS and large black hole phases \cite{Hawking:1982dh} and the van der Waals-type phase transition in charged AdS black holes \cite{Kubiznak:2012}. These discoveries reveal that black holes possess thermodynamic behavior as rich as that of ordinary matter and are intimately connected to key phenomena in condensed-matter physics through the AdS/conformal field theory (CFT) correspondence.

Recent work has increasingly explored these phase transitions from geometric and topological perspectives, providing new insight into this important field. It has been demonstrated that the synchronized multivalued behavior of dynamical quantities such as the Lyapunov exponent and geometric quantities such as the intrinsic or extrinsic curvature near the phase transition point can diagnose a first-order phase transition \cite{Guo:2022,Yang:2023,Lyu:2024,Kumara:2024,Du:2025,Shukla:2024,Gogoi:2024,Chen:2025xqc,R:2025gok,Awal:2025irl,Yang:2025fvm,Kumar:2025kzt,Guo:2025pit,Bezboruah:2025udi,Ali:2025ooh,Zhang:2025,Xie:2025auj,Zhang:2025kqd}. Moreover, since the Lyapunov exponent is quantitatively linked to geometric quantities in both null and timelike cases, this connection has established a unified perspective linking dynamics, spacetime geometry, and first-order phase transitions \cite{Gallo:2025,Zhang:2025,Zhang:2025kqd}. Separately, topological classification schemes have revealed that black holes can be categorized according to their topological properties \cite{Wei:2021vdx,Wei:2022dzw,Wei:2024gfz}, offering new insights for a deeper understanding of the fundamental nature of quantum gravity.

However, the mathematical and physical mechanism linking multivaluedness to first-order phase transitions remains unclear. A central puzzle persists: Is multivaluedness merely an accidental phenomenon, or is it an intrinsic mathematical signature of the first-order phase transition itself? Furthermore, while topological classification provides a powerful global perspective, a consequent question arises: Can we develop a criterion based on local geometric features that can independently diagnose the phase transition while also offering a complementary perspective to the global topological approach?

In this paper, we answer both questions by constructing a geometric framework that integrates real analysis and covering space theory. We prove that if and only if the temperature $T$, expressed as a function of the horizon radius $r_+$, has two nondegenerate critical points, the inverse mapping $r_+(T)$ necessarily becomes three-valued within the spinodal region. This result means that a single temperature corresponds to three distinct horizon radii, representing the large, intermediate, and small black hole solutions. For a geometric perspective, this phenomenon originates from the fact that these nondegenerate critical points act as fold singularities of the temperature mapping. They induce a branched three-sheeted structure in the thermodynamic parameter space, which provides a universal, intrinsic geometric origin for the synchronized multivalued behavior exhibited by all physical and geometric quantities during a first-order phase transition.

Based on these considerations, we propose a universal geometric criterion for small/large black hole first-order phase transitions: examining whether the curve of the temperature function $T(r_+)$ possesses two extrema (a local maximum and a local minimum). Equivalently, this criterion requires that $\partial T/{\partial r_+} = 0$ admits two distinct positive real roots. This criterion provides an intuitive and computable diagnostic tool that enables a clear classification based on the local geometric structure underlying the phase transition. Our work demonstrates that the spacetime structure governed by the Einstein equations induces specific geometric features, such as nondegenerate critical points in the thermodynamic parameter space of a black hole. This results in quantities such as the photon sphere radius, Lyapunov exponent, and intrinsic/extrinsic curvatures exhibiting multivalued or non-monotonic dependence on temperature $T$ \cite{Guo:2022,Yang:2023,Lyu:2024,Kumara:2024,Du:2025,Shukla:2024,Gogoi:2024,Chen:2025xqc,R:2025gok,Awal:2025irl,Yang:2025fvm,Kumar:2025kzt,Guo:2025pit,Bezboruah:2025udi,Ali:2025ooh,Zhang:2025,Xie:2025auj,Zhang:2025kqd,PS01,PS02,PS03,PS04,PS05,PS06,PS07,PS08,PS09}. These features intrinsically encode the information of phase transitions. Therefore, the geometry of the thermodynamic parameter space offers a profound and unified perspective for understanding black hole phase transitions. We set $G=c=k_B=\hbar=1$ in this paper.

\section{Black hole first-order phase transitions}

In this paper, we focus on first-order phase transitions of black holes, beginning with a review of the relevant thermodynamic relations. We consider a $(3+1)$-dimensional spherically symmetric black hole described by the metric
\begin{align}
ds^{2}=-fdt^{2}+\frac{1}{f}dr^{2}+r^{2}d\Omega^{2},
\end{align}
where $f$ is a function of the radial coordinate $r$, and $d\Omega^{2}$ is the unit 2-sphere. The Hawking temperature is given by
\begin{align}
T = \frac{f'}{4\pi} \bigg|_{r_+},
\end{align}
with $r_+$ being the horizon radius defined by $f(r_+)=0$ and $f'=df/dr$. The Gibbs free energy reads
\begin{align} 
F = M-TS,
\end{align}
where $M$ is the ADM mass of the black hole and $S =\pi r_+^2$ is the entropy. The critical condition for a phase transition is
\begin{align}
\frac{\partial T}{\partial r_+} = \frac{\partial^2 T}{\partial r_+^2} = 0.\label{c}
\end{align}
When this condition is met, the black hole undergoes a first-order phase transition, and the $F(T)$ curve displays a characteristic swallowtail structure.

Considering an unstable null circular orbit (light ring) at radius $r_{LR}$ near the black hole, its Gaussian curvature is given by \cite{Gallo:2025}
\begin{align}
K(r_{LR}) = \left[ \frac{f}{2} \left( f'' - \frac{f'}{r} \right) \right] \Bigg|_{r_{LR}},
\end{align}
where $r_{LR}$ is determined by solving
\begin{align}
2f(r_{LR})=r_{LR}f'(r_{LR}).
\end{align}

We take the Reissner–Nordström–anti-de Sitter (RN–AdS) black hole as an example. Its metric function is given by
\begin{align}
f(r)= 1 - \frac{2M}{r} + \frac{Q^2}{r^2} + \frac{r^2}{\ell^2},\label{fr}
\end{align}
where $Q$ is the electric charge and $\ell$ is the AdS radius. We introduce the following scaling
\begin{align}
\tilde{r}_+ = \frac{r_+}{\ell}, \quad \tilde{Q} = \frac{Q}{\ell}, \quad \tilde{M} = \frac{M}{\ell}, \quad \tilde{F} = \frac{F}{\ell}, \quad \tilde{T} = T\ell.
\end{align}

\begin{figure}
\includegraphics[width=0.46\textwidth, height=0.23\textheight]{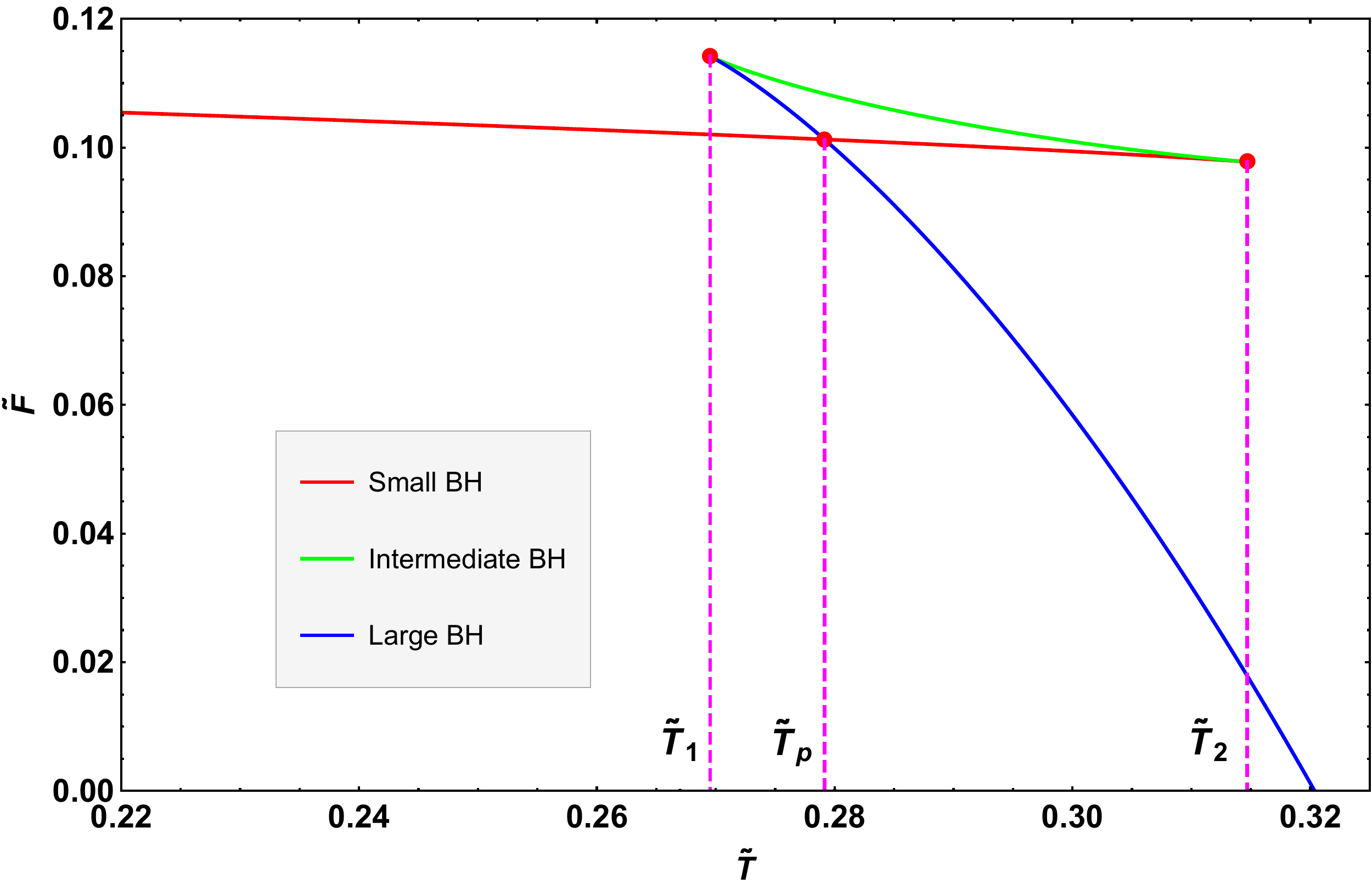} \hspace{1cm}
\includegraphics[width=0.46\textwidth, height=0.23\textheight]{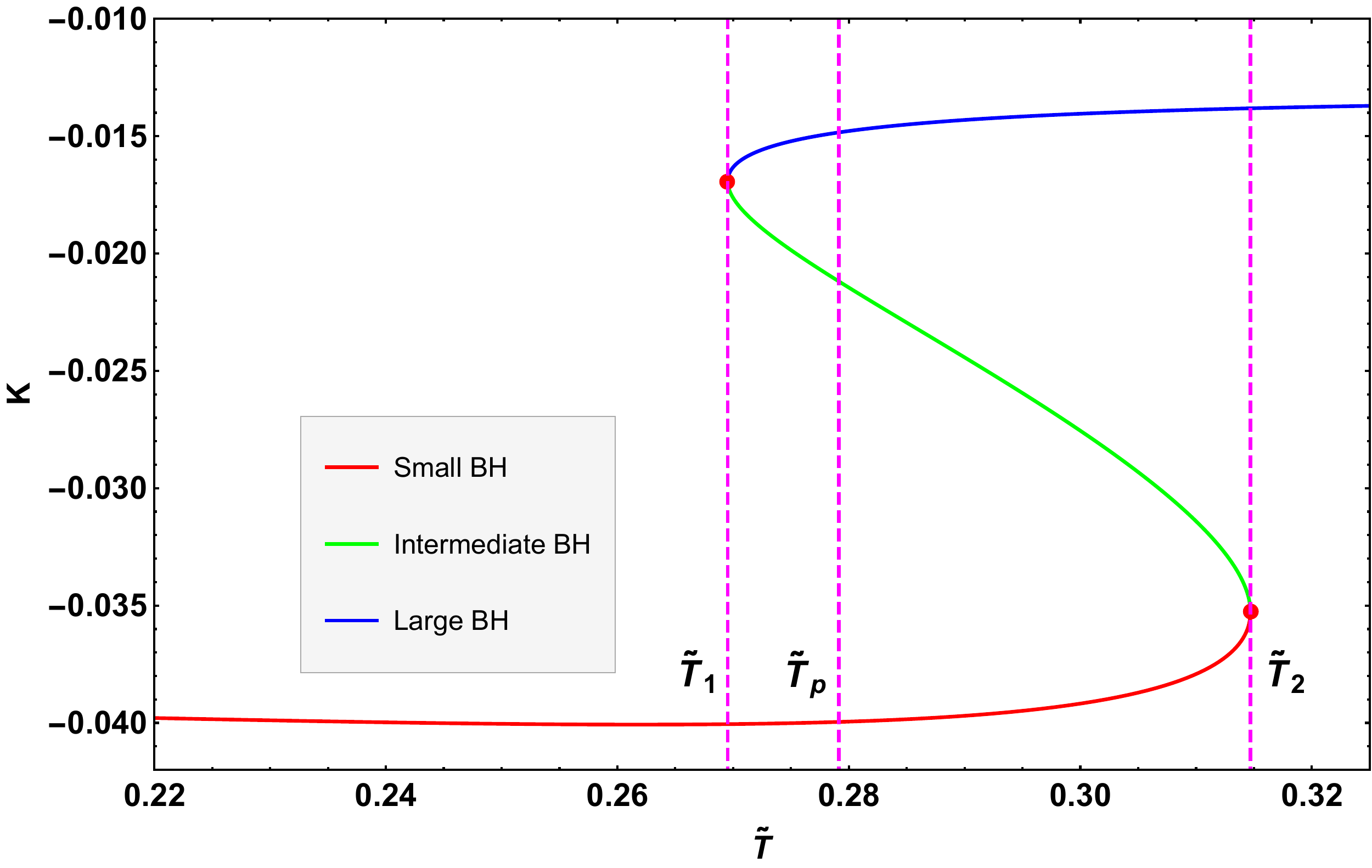}
\caption{\label{1} Thermodynamic and geometric signatures of the first-order phase transition in RN–AdS black holes. \emph{Left}: Free energy $\tilde{F}$ versus temperature $\tilde{T}$ for RN–AdS black holes. \emph{Right}: Gaussian curvature $K$ of unstable null orbits versus temperature $\tilde{T}$ for RN–AdS black holes. $\tilde{Q}=\frac{1}{8.66},~\tilde{Q}_c=\frac{1}{6},~\tilde{Q}<\tilde{Q}_c$. The synchronized multivalued behavior of $K$ in the spinodal region $\tilde{T} \in (\tilde{T}_{1},\, \tilde{T}_{2})$ corresponds to the swallowtail structure in the free energy, with the phase transition occurring at $\tilde{T}_p$.}
\end{figure}

As shown in Fig.~\ref{1}, for the RN–AdS black hole undergoing a first-order phase transition, the curves of $K(\tilde{T})$ and $\tilde{F}(\tilde{T})$ exhibit synchronized multivalued behavior inside the spinodal region \cite{Zhang:2025}, demonstrating that such multivaluedness signals the occurrence of the phase transition. The critical charge $\tilde{Q}_c=1/6$ is given by Eq.~(\ref{c}). At the two endpoints $\tilde{T}_1$ and $\tilde{T}_2$ of the spinodal region, the thermodynamic condition $\partial T/ \partial r_+ =0$ holds. Moreover, because the Gaussian curvature $K$ and the Lyapunov exponent $\lambda$ that characterize orbital chaos are related by $K=-\lambda^2$ at the light ring $r_{LR}$ \cite{Gallo:2025}, the $\lambda(\tilde{T})$ curve consequently also displays multivaluedness in the spinodal region \cite{Guo:2022}. In addition, an analogous correspondence holds for unstable timelike circular orbits \cite{Zhang:2025kqd}.

The multivalued behavior of these quantities, which reflect chaotic dynamics and spacetime geometry, shows a precise consistency with the thermodynamic free energy $\tilde{F}$. This indicates that phase transition information is universally encoded in multiple properties of spacetime \cite{Zhang:2025,Zhang:2025kqd}. In the following analysis, we will investigate this characteristic phenomenon and uncover the deep mathematical structure and physical mechanisms behind it.

As shown in Fig.~\ref{1}, for the RN–AdS black hole undergoing a first-order phase transition, the curves of $K(\tilde{T})$ and $\tilde{F}(\tilde{T})$ exhibit synchronized multivalued behavior inside the spinodal region \cite{Zhang:2025}, demonstrating that such multivaluedness signals the occurrence of the phase transition. The critical charge $\tilde{Q}_c=1/6$ is given by Eq.~(\ref{c}). At the two endpoints $\tilde{T}_1$ and $\tilde{T}_2$ of the spinodal region, the thermodynamic condition $\partial T/ \partial r_+ =0$ holds. Moreover, because the Gaussian curvature $K$ and the Lyapunov exponent $\lambda$ that characterizes orbital chaos are related by $K=-\lambda^2$ at the light ring $r_{LR}$ \cite{Gallo:2025}, the $\lambda(\tilde{T})$ curve consequently also displays multivaluedness in the spinodal region \cite{Guo:2022}. In addition, an analogous correspondence holds for unstable timelike circular orbits \cite{Zhang:2025kqd}.

The multivalued behavior of these quantities, which reflect chaotic dynamics and spacetime geometry, shows a precise consistency with the thermodynamic free energy $\tilde{F}$. This indicates that phase transition information is universally encoded in multiple properties of spacetime \cite{Zhang:2025,Zhang:2025kqd}. In the following analysis, we will investigate this characteristic phenomenon and uncover the deep mathematical structure and physical mechanisms behind it.

\section{The mathematical origin of multivaluedness}

To investigate the universal origin of the multivalued phenomenon described above, we constructed a geometric framework that combines real analysis and covering space theory. The validity of this framework is based on the essential requirements obtained from a first‑order phase transition: the temperature $T(r_+)$ as a function of the horizon radius is sufficiently smooth. When a first-order phase transition occurs, $T(r_+)$ has two nondegenerate critical points, denoted $T_1$ and $T_2$. Furthermore, $T(r_+)$ is strictly monotonic outside the spinodal interval. Moreover, $T(r_+)$ is not a constant function over its entire domain.

Within this geometric framework, we prove a key result: the conditions stated above necessarily lead $r_+(T)$ to become three‑valued inside the spinodal region. Moreover, this multivaluedness is inherited by any strictly monotonic physical or geometric quantity $\mathcal{F}(r_+)$: if $\mathcal{F}(r_+)$ is strictly monotonic, then the multivaluedness of $T(r_+)$ implies that $\mathcal{F}(T)$ is also multivalued. The proof proceeds as follows. Analyzing the monotonic intervals and applying the intermediate value theorem and Darboux’s theorem, we can rigorously construct three continuous solution branches, corresponding to large, intermediate, and small black hole solutions. Consequently, $r_+(T)$ is three‑valued in the open interval $(T_1,\,T_2)$. At the endpoints $T_1$ and $T_2$, where the implicit function theorem fails, $r_+(T)$ is two-valued (with a double root). Thus, multivaluedness persists on the closed interval $\mathcal{S}=[T_1,\,T_2]$.

This conclusion has a deeper geometric counterpart in the theory of covering spaces. At each critical point $r_v$ (with $v=1,2$), the Morse lemma guarantees that $T(r_+)$ can be locally expressed as $T \approx T_v - y^2$ for a local maximum and $T \approx T_v + y^2$ for a local minimum, where $y$ measures the deviation from the nondegenerate critical point \cite{Morse:1925}. These points are therefore fold singularities, which directly implies that near $T_1$ or $T_2$, a single temperature corresponds to two distinct horizon radii $r_+$ (one double root and one distinct root). This folding of the parameter space is formalized by introducing a covering manifold. We define a one-dimensional manifold $\mathcal{M}$ parameterized by all horizon radii $r_+\in I\subset \mathbb{R}$ that correspond to such black hole solutions. We accordingly construct a branched three-sheeted covering manifold $\widetilde{\mathcal{M}} = J_1 \sqcup J_2 \sqcup J_3$, where $J_1 = (\mathcal{A},\,r_1),~J_2 = (r_1,\,r_2),~J_3 = (r_2,\,\mathcal{B})$ with $\mathcal{A} < r_1 < r_2 < \mathcal{B}$. Here, $\mathcal{A}$ and $\mathcal{B}$ are points in the domain $I \subset \mathbb{R}^+$, and the interval $(\mathcal{A},\,\mathcal{B})$ constitutes the relevant parameter range for the phase transition. The projection $\widetilde{\pi} : \widetilde{\mathcal{M}} \rightarrow \mathbb{R}$, defined by $\widetilde{\pi}(r_+) = T(r_+)$, is injective on each branch. The topological structure of the base manifold $\mathcal{M}$ ensures that its covering space $\widetilde{\mathcal{M}}$ must be three-sheeted on the spinodal region. This explains the geometric origin of multivaluedness within the black hole parameter manifold.

Consequently, any physical quantity $\mathcal{F}$ that depends on $r_+$ inherits this multivaluedness. Explicitly, $\mathcal{F}(T)$ can be regarded as the composition of its lift $\mathcal{F}^\uparrow$ on the covering space with the inverse projection $\widetilde{\pi}^{-1}(T)$. Since $\widetilde{\pi}^{-1}(T)$ yields three distinct $r_+$ values and $\mathcal{F}^\uparrow$ is not identically equal on different branches, $\mathcal{F}(T)$ naturally exhibits multivaluedness. Conversely, observing multivaluedness in $\mathcal{F}(T)$ implies the existence of nondegenerate critical points in $T(r_+)$, thereby diagnosing a first‑order phase transition. This explains why the use of multivaluedness to detect first‑order phase transitions in Refs.~\cite{Guo:2022,Yang:2023,Lyu:2024,Kumara:2024,Du:2025,Shukla:2024,Gogoi:2024,Chen:2025xqc,R:2025gok,Awal:2025irl,Yang:2025fvm,Kumar:2025kzt,Guo:2025pit,Bezboruah:2025udi,Ali:2025ooh,Zhang:2025,Xie:2025auj,Zhang:2025kqd} is justified, thereby providing a rigorous mathematical foundation for this method. Complete technical details are given in the Appendix.

\section{A local geometric criterion for classifying phase transitions}
Since $\mathcal{F}(T)$ inherits the multivaluedness of $r_+(T)$ during a first-order phase transition, we propose a classification scheme for black holes based on the local geometric criterion introduced above. It categorizes black holes into classes $A1$, $A2$, and $B$ according to the number of local extrema in the $T(r_+)$ curve, as shown in Table~\ref{tab1}. The $A2$ class corresponds to curves with two distinct extrema, signaling two nondegenerate critical points. This leads to a three-valued $\mathcal{F}(T)$, resulting in a first-order phase transition. In contrast, the $A1$ class exhibits only one critical point and therefore does not display the characteristic features of a first-order phase transition. Finally, for the $B$ class, the curve has no local extrema, thus $\mathcal{F}(T)$ is strictly monotonic, which means the absence of a phase transition for such black holes.

\begin{table}[htbp]
\centering
\setlength{\tabcolsep}{3.2pt}
\begin{tabular*}{\textwidth}{c@{\extracolsep{\fill}}lcccccc}
\hline
  & Extrema & First-order phase transition & Branch\\
\hline
\(A1\) & 1 & no & 2 \\
\(A2\) & 2 & yes & 3 \\
\(B\)  & no & no & 1 \\
\hline
\end{tabular*}
\caption{Black hole classification based on the number of extrema of the $T(r_+)$ curve.}\label{tab1}
\end{table}

\begin{figure}
\includegraphics[width=0.46\textwidth, height=0.23\textheight]{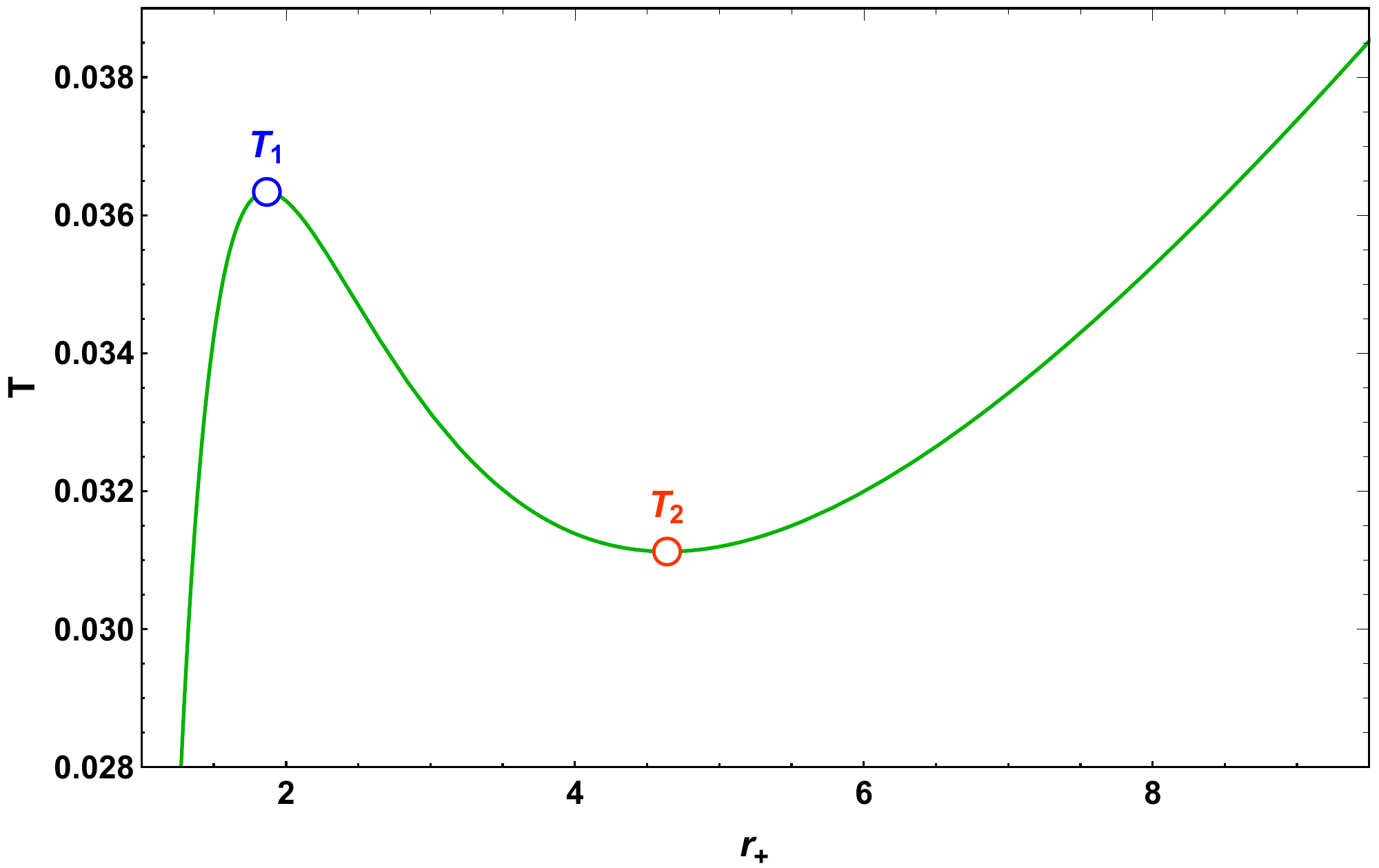} \hspace{1cm}
\includegraphics[width=0.46\textwidth, height=0.23\textheight]{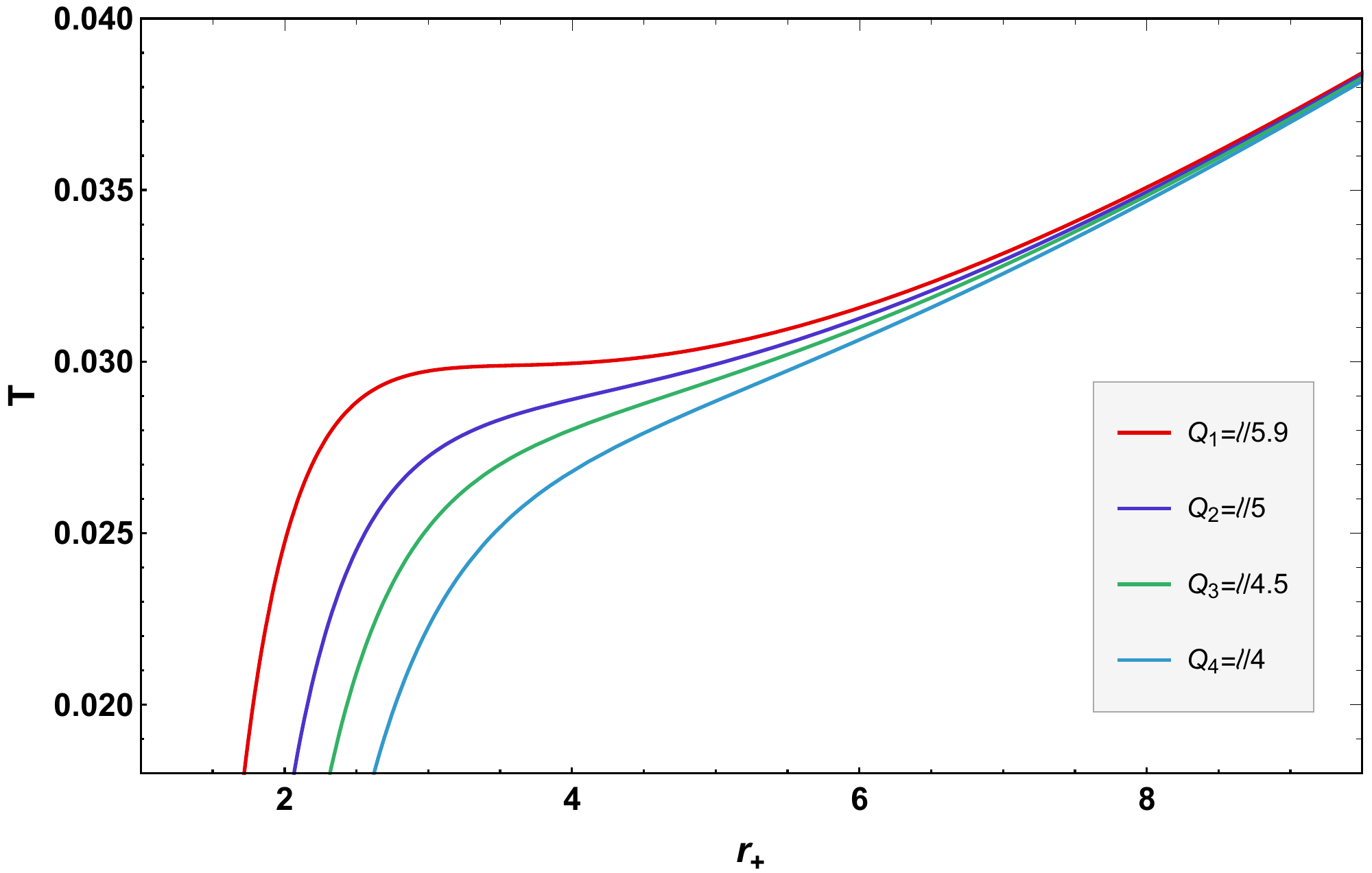}
\caption{\label{2} Temperature curves $T(r_+)$ for the RN–AdS black hole with and without a first‑order phase transition. \emph{Left}: For $Q=(\ell /8.66)<Q_c$, a first‑order phase transition occurs, and the $T(r_+)$ curve exhibits two local extrema ($T_1$ and $T_2$), signaling multivaluedness. \emph{Right}: For $Q>Q_c$ ($Q_1=\ell /5.9,~Q_2=\ell /5,~Q_3=\ell /4.5,~Q_4=\ell /4$ with $\ell=8.66$), the curve shows no local extremum, indicating the absence of a first‑order phase transition.}
\end{figure}

We illustrate this criterion with the RN–AdS black holes. For the RN‑AdS black hole, the metric function is Eq.~(\ref{fr}), and the Hawking temperature reads
\begin{align}
T =\frac{1}{4\pi} \left(\frac{1}{r_+} - \frac{Q^2}{r_+^3} + \frac{3r_+}{\ell^2}\right).
\end{align}
As shown in Fig.~\ref{2}, for an RN–AdS black hole with $Q=(\ell /8.66)<Q_c$, the $T(r_+)$ curve displays a local maximum $T_1$ and a local minimum $T_2$, signaling a typical first-order phase transition. In contrast, for a series of charges $Q > Q_c$ (such as $\ell/5.9$, $\ell/5$, $\ell/4.5$, $\ell/4$), the curves are monotonic with no local extrema, indicating the absence of a first-order phase transition. This result agrees with the conclusion obtained from the thermodynamic analysis \cite{Guo:2022}, thus providing support for our geometric framework.

This classification is equivalent to checking whether $\partial T/ \partial r_+ =0$ admits two distinct positive real roots. We illustrate it using the RN‑AdS and Schwarzschild‑AdS (SAdS) black holes. Setting $T'(r_+) =0$ yields
\begin{align}
-r_+^2 + 3Q^2 + \frac{3r_+^4}{\ell^2} = 0.
\end{align}
The discriminant of this quadratic equation is $\Delta = 1 - \frac{36Q^2}{\ell^2} > 0$. When $\Delta>0$ (equivalently $\ell>6Q$), the equation has two distinct positive real roots, indicating a first-order phase transition and placing the black hole in the
$A2$ class. The condition $\ell > 6Q$ matches the critical charge from the standard thermodynamic criticality condition Eq.~(\ref{c}), which confirms the effectiveness of our multivaluedness‑based diagnostic.

For the SAdS black hole, setting $Q=0$ leads
\begin{align}
T =\frac{1}{4\pi} \left( \frac{1}{r_+} + \frac{3r_+}{\ell^2}\right).
\end{align}
Solving $T'(r_+) =0$ yields
\begin{align}
r_+ = \pm \sqrt{\frac{\ell^2}{3}}.
\end{align}
The only physical result is $r_+ = \sqrt{\ell^2/{3}}$. Hence, the $T(r_+)$ curve has only one critical point. The SAdS black hole therefore does not exhibit the characteristics of a first-order phase transition and belongs to the $A1$ class.

Our classification scheme, rooted in the analysis of local geometry in the thermodynamic parameter space, offers a complementary perspective to the framework in Ref.~\cite{Wei:2024gfz}, which is based on global topological invariants. Their work assigns the RN–AdS and SAdS black holes to the topological classes $W^{0+}$ and $W^{0-}$, respectively. These two distinct perspectives highlight the depth and effectiveness of the geometric approach in studying black hole thermodynamics.

\section{Extensions and universality of the geometric framework}

The geometric mechanism proposed in this paper is not restricted to spherically symmetric black holes. For rotating black holes such as Kerr‑AdS and Kerr‑Newman‑AdS, the temperature curve $T(r_+)$ at fixed angular momentum (and charge) also exhibits two local extrema when a first‑order phase transition occurs \cite{Chen:2025xqc,Yang:2025fvm}. According to our criterion, these rotating black holes therefore belong to the $A2$ class, and the multivaluedness of related dynamical or geometric quantities arises from the same folded geometry.

Furthermore, in Refs.~\cite{PS01,PS02,PS03,PS04,PS05,PS06,PS07,PS08,PS09}, diagnosing first-order phase transitions by observing non-monotonic (multivalued) behavior in the diagram of reduced temperature $T/T_c$ versus reduced photon sphere radius $r_{ps}/r_{psc}$ (where $T_c$ is the critical temperature and $r_{psc}$ the critical photon sphere radius) is essentially equivalent to checking whether $T(r_+)$ possesses two nondegenerate critical points. This equivalence holds because the photon sphere radius $r_{ps}$ is a physical quantity that continuously depends on the horizon radius $r_+$, and its values do not coincide in different branches. In most known black hole solutions, this dependence is in fact monotonic, which yields a one-to-one correspondence between the branches. Consequently, by taking $\mathcal{F}=r_{ps}$, the multivaluedness of $T(r_+)$ directly implies that of $T(r_{ps})$, and scaling these physical quantities does not change this fundamental property.

\section{Further comments and conclusions}

This work reveals the universal mathematical mechanism behind the multivalued behavior in the first-order phase transitions of spherically symmetric or rotating black holes. By constructing a unified geometric framework that integrates real analysis and covering space theory, we rigorously prove that the multivaluedness of any physical or geometric quantity $\mathcal{F}(T)$ during a first-order phase transition is not incidental but a topological necessity, originating from the presence of two nondegenerate critical points in the temperature function $T(r_+)$. 

This geometric mechanism thus provides a unified foundation for diagnosing phase transitions via the multivaluedness of different physical quantities. Specifically, this finding clarifies the universal geometric origin of the synchronized multivalued behavior observed in the dynamical Lyapunov exponent and the geometric curvatures, and links it to the first-order phase transition signaled by the swallowtail structure. Moreover, our framework successfully incorporates a phase transition probe based on photon spheres. This shows that they are all rooted in the topological structure of the thermodynamic parameter space. Furthermore, the multivaluedness probes that arise from different perspectives remain highly valuable for understanding the distinctive physical and geometric features of first-order phase transitions in various gravitational theories. Thus, our framework demonstrates the effectiveness of these diagnostic methods from the viewpoint of covering theory and establishes the role of multivaluedness as a key bridge connecting thermodynamics, dynamics, and geometry.

As a direct application of this framework, we propose the following geometric criterion for diagnosing a phase transition: a black hole undergoes a small/large black hole first-order phase transition if and only if the temperature curve $T(r_+)$ exhibits two local extrema. This criterion is equivalent to testing whether $\partial T/ \partial r_+ =0$ has two distinct positive real roots, providing a universal and intuitive tool. Furthermore, we propose a classification scheme that classifies black holes as $A1$, $A2$, and $B$ according to the number of nondegenerate critical points. This classification, based on local geometric properties, forms a complementary perspective to the scheme presented in Ref.~\cite{Wei:2024gfz}, which relies on global topological invariants. 

It is worth noting that the local classification scheme can be naturally generalized to the case where the temperature function possesses multiple extremal points, thereby corresponding precisely to the global and complex analytic classification schemes \cite{Zhang:2026sht}. Moreover, since the essence of multivaluedness is the Maxwell equal‑area law, the local framework can be extended to other ensembles.

In summary, our results demonstrate that spacetime geometry profoundly encodes thermodynamic information. The geometric criterion we propose unifies diverse perspectives within a single theoretical framework. This opens new pathways for exploring the interconnections among thermodynamics, dynamics, and geometry in more complex gravitational systems, such as higher-dimensional black holes or black holes in modified theories of gravity. By constructing a concrete geometric framework linking these three areas, this work provides a unified theoretical foundation for understanding their intrinsic relationships during black hole phase transitions.

\section*{Acknowledgments}
This work was supported by the National Natural Science Foundation of China (Grant Nos. 12533001, 12473001, 12575049), the National SKA Program of China (Grant Nos. 2022SKA0110200, 2022SKA0110203), the China Manned Space Program (Grant No. CMS-CSST-2025-A02), and the 111 Project (Grant No. B16009).

\section*{Data availability statement}
This manuscript has no associated data.
[Author’s comment: Data sharing not applicable to this article as no
datasets were generated or analysed during the current study].

\section*{Code availability statement}
This manuscript has no associated code/
software. [Author’s comment: Code/Software sharing not applicable to
this article as no code/software was generated or analysed during the
current study].

\bibliography{main}

\appendix

\section{Physical nature and mathematical conditions}\label{appA}

The mathematical conditions we introduce are grounded in basic physical facts of a first-order phase transition. First, the temperature as a function $T(r_+)$ is sufficiently smooth, i.e., $T \in C^p(I)$ with $p \ge 2$ and $I \subset \mathbb{R}$. More importantly, during a first-order phase transition, this function exhibits two nondegenerate critical points, namely $T'(r_v) =0$ and $T''(r_v) \ne 0$ for $v = 1,2$, which correspond to the two endpoints $T_1$ and $T_2$ of the spinodal region. This is because, when a first-order phase transition occurs, for Maxwell's equal area law to be meaningful, the function must have two local extremum points with non-zero second derivatives. Hence, for a single-variable function, such extremum points are precisely nondegenerate critical points. Outside the spinodal region, $T(r_+)$ is strictly monotonic. Finally, $T(r_+)$ is not a constant function on its entire domain. For a black hole that satisfies the Einstein equations, its Hawking temperature $T(r_+)$ may inherently satisfy or violate these conditions, which directly determines its phase transition behavior.

\section{Analytic perspective}\label{appendixB}

Let the domain of $T(r_+)$ be an interval $I \subset \mathbb{R}^+$. For the analysis of the phase transition, we focus on a subinterval $(\mathcal{A},\,\mathcal{B}) \subset I$ containing the critical points, with $\mathcal{A} < r_1 < r_2 < \mathcal{B}$. We first examine the behavior of the function $T(r_+)$ near its critical points from an analytic viewpoint. Let $T_1=T(r_1)$ and $T_2=T(r_2)$. According to Darboux's theorem, the derivative $\partial T/ \partial r_+$ does not change sign on the interval $(r_1,\,r_2)$, otherwise there would exist a point $r_\chi \in V,~V = (r_1,\,r_2)$ such that $\left( \partial T/\partial r_+ \right) \big|_{r_\chi} = 0$, contradicting the assumption that $r_1$ and $r_2$ are the only critical points. We assume $\left( \partial T/\partial r_+ \right) \big|_{r_0} < 0$ for all $r_0 \in V$, which implies that $T_1$ is a local maximum and $T_2$ a local minimum. The proof proceeds identically if the sign is reversed.

Because $T(r_+)$ is continuously differentiable at $r_1$, there exists a point $a \in I$ with $\mathcal{A}<a<r_1$ such that $T(r_+)$ is monotonic on $(a,\,r_1)$. Since $r_1$ is a local extremum, we have $\left( \partial T/\partial r_+ \right) \big|_{r_a} > 0$. We define the function on this interval as Branch1 (the small black hole branch). On the interval $(r_1,\,r_2)$, because $\left( \partial T/\partial r_+ \right) \big|_{r_0} < 0$, we define the function as Branch2 (the intermediate black hole branch). Similarly, there exists a point $b \in I$ with $\mathcal{B}>b>r_2$ such that $T(r_+)$ is monotonic on $(r_2,\,b)$ and $\left( \partial T/\partial r_+ \right) \big|_{r_b} > 0$. The function on this interval is defined as Branch3 (the large black hole branch).

It can be shown that for every $T_M \in (T_1,\,T_2)$, $T(r_+)=T_M$ has one root on each of the three branches. On Branch1, since $T(r_+)$ is monotonically increasing on $(a,\,r_1)$, we have $\lim\limits_{r_+ \to r_1} T(r_+) < T_1$. Given $T_M<T_1$, the intermediate value theorem guarantees a unique $r_+^1 \in (a,\,r_1)$ such that $T(r_+^1) = T_M$. Similarly, on Branch2, with the condition $T_1<T_M<T_2$, the intermediate value theorem yields a unique $r_+^2 \in V$ satisfying $T(r_+^2) = T_M$. On Branch3, because $T(r_+)$ is monotonically increasing on $(r_2,\,b)$, we have $\lim\limits_{r_+ \to r_2} T(r_+) > T_2$. Then, by the intermediate value theorem there exists a unique $r_+^3 \in (r_2,\,b)$ for which $T(r_+^3) = T_M$. Since $r_+^1 \in (a,\,r_1),~r_+^2\in (r_1,\,r_2)$ and $r_+^3 \in (r_2,\,b)$, the three values $r_+^1,~r_+^2$ and $r_+^3$ are distinct positive real numbers. Hence, the multivaluedness of the inverse function $r_+(T)$ over the temperature interval $(T_1,\,T_2)$ is inevitable.

We now demonstrate how the multivaluedness of any physical or geometric quantity $\mathcal{F}$ in the interval $\mathcal{S}=[T_1,\,T_2]$ is inherited from the multivaluedness of $r_+(T)$ during a black hole first‑order phase transition. Let $\mathcal{F}: I \to \mathbb{R}$ be a continuous function of the horizon radius, denoted as $\mathcal{F}(r_+)$. This requires that $r_+(T)$ itself be multivalued on $(T_1, T_2)$, having at least two distinct branches $r_i(T)$ and $r_j(T)$ with $r_i(T) \neq r_j(T)$. Additionally, $\mathcal{F}$ must not be a constant function and satisfy
\begin{equation}
\mathcal{F} \circ r_i \neq \mathcal{F} \circ r_j,
\end{equation}
on that interval, meaning $\mathcal{F}$ is not identically equal on the two branches. When these conditions are met, $\mathcal{F}(T)$ itself becomes multivalued. More precisely, there exists a non-empty open subinterval $J \subset \mathcal{S}$ such that
\begin{equation}
\mathcal{F}[r_i(T)] \neq \mathcal{F}[r_j(T)],~\forall T \in J.
\end{equation}
The condition above serves to exclude physical quantities that can be expressed as single-valued functions of the temperature alone (for example, $\mathcal{F}=T^2$). Although such quantities also depend indirectly on $r_+$, the value of $\mathcal{F}$ is only determined by $T$, regardless of which branch $r_+(T)$ is chosen. Hence, the $\mathcal{F}-T$ relation remains a single‑valued (monotonic) function and does not inherit the multivalued structure from $r_+(T)$. We are instead interested in quantities that genuinely distinguish different geometric states, characterized by an explicit dependence on $r_+$ that yields distinct values on different branches and thus an implicit relation between $\mathcal{F}$ and $T$ via $r_+$.

We now prove the multivaluedness of $\mathcal{F}(T)$ on the subinterval 
$J$. Because $\mathcal{F} \circ r_i$ and $\mathcal{F} \circ r_j$ are continuous on $(T_1,\,T_2)$ and are not identically equal, there exists a temperature $T_0 \in (T_1,\,T_2)$ such that
\begin{equation}
\mathcal{F}[r_i(T_0)] \neq \mathcal{F}[r_j(T_0)].
\end{equation}
Define the difference function
\begin{equation}
D(T) = \mathcal{F}[r_i(T)] - \mathcal{F}[r_j(T)],
\end{equation}
Since $\mathcal{F}$, $r_i$ and $r_j$ are continuous, $D(T)$ is continuous on $(T_1,\,T_2)$, and satisfies $D(T_0)\ne 0$. By the continuity of $D$, there exists an interval $J$ containing $T_0$ such that $D(T)\ne 0$ for all $T \in J$. Hence, on the interval $J$, $\mathcal{F}(T)$ takes at least two distinct values, establishing the multivaluedness of $\mathcal{F}(T)$ on that subinterval $J$.

The condition that $\mathcal{F}[r_i(T)]$ and $\mathcal{F}[r_j(T)]$ are not identically equal leads to a stronger result: the multivaluedness persists everywhere on $(T_1,\,T_2)$ except possibly on a negligible set $Z$. For completeness, we must still show that $\mathcal{F}(T)$ is multivalued on the whole interval $\mathcal{S}$. To understand this, define this zero set as
\begin{equation}
Z = \{ T \in [T_1,\,T_2] \mid D(T) = 0 \}.
\end{equation}
Note that the set $Z$ is the zero set of the function $D(T)$ on $\mathcal{S}$. Since $D(T)$ is continuous on $\mathcal{S}$, $Z$ is a closed set. Because $D(T)$ is not identically zero, when $Z \ne \mathcal{S}$ its complement $\mathcal{S} \setminus Z$ is dense in $\mathcal{S}$, and $D(T)\ne 0$ for every $T\in (T_1,\,T_2) \setminus Z$.

Furthermore, since the thermodynamic conditions $T'(r_v) =0$ and $T''(r_v) \ne 0$ at the endpoints $T_1$ and $T_2$, the implicit function theorem fails there, which also forces $\mathcal{F}(T)$ to be multivalued at these boundary points. Consequently, $\mathcal{F}(T)$ exhibits multivaluedness on the whole closed interval $[T_1,\,T_2]$, except possibly on the measure‑zero set $Z$ within the open interval.

\section{Covering space perspective}\label{appendixC}
The analytic conclusion described above has a natural counterpart from a geometric perspective. Consider the thermodynamic system such as black holes, where each black hole with fixed physical parameters (such as mass $M$, charge $Q$, angular momentum $J$) has a definite horizon radius $r_+$. We define a manifold $\mathcal{M}$ parameterized by all horizon radii $r_+\in I\subset \mathbb{R}$ that correspond to such black hole solutions. The temperature $T$ is a smooth function on $\mathcal{M}$, which is equivalent to the existence of a map
\begin{equation}
\pi \colon \mathcal{M} \to \mathbb{R},~\pi(r_+) = T(r_+).
\end{equation}
This work focuses on the thermodynamics of non-extremal black holes ($T>0$). For mathematical convenience, we take the codomain of the temperature map to be $\mathbb{R}$. Under the assumption of a first-order phase transition, we require the temperature map $\pi$ to be smooth (at least $C^2$). Furthermore, the map must possess two nondegenerate critical points $r_1, r_2 \in \mathcal{M}$ (with $r_1 < r_2$), satisfying
\begin{equation}
\left(\frac{\partial T}{\partial r_+}\right) \bigg|_{r_1} = \left(\frac{\partial T}{\partial r_+}\right) \bigg|_{r_2} = 0,~\left(\frac{\partial^2 T}{\partial r_+^2}\right) \bigg|_{r_1} \ne 0, ~\left(\frac{\partial^2 T}{\partial r_+^2}\right) \bigg|_{r_2} \ne 0.
\end{equation}
These conditions imply that $r_1$ and $r_2$ are nondegenerate critical points of the function $T$ on $\mathcal{M}$. Moreover, there exist intervals $(\mathcal{A},\,r_1),~(r_1,\,r_2)$ and $(r_2,\,\mathcal{B})$, on these intervals, $T(r_+)$ is strictly monotonic. Finally, $T(r_+)$ is not constant on the spinodal region $\mathcal{S}=[T_1,\,T_2]$.

We again assume that $T(r_+)$ has a local maximum $T_1$ at $r_1$ and a local minimum $T_2$ at $r_2$ (the proof is identical if the extrema types are reversed). The Morse lemma ensures that near a nondegenerate critical point $p$, there exist local coordinates $(y^1,\,\cdots,\,y^n)$ such that the function can be written in the standard quadratic form
\begin{equation}
f(y) = f(p) - (y^1)^2 - \cdots - (y^k)^2 + (y^{k+1})^2 + \cdots + (y^n)^2,
\end{equation}
where $y$ represents the local coordinate that measures the deviation from the critical point, and $k$ is the index, or the number of negative eigenvalues of the Hessian matrix. Because the critical points are nondegenerate, the Morse lemma ensures the existence of such a local coordinate $y$, which is related to $r_+$ by a diffeomorphism.

Since $r_1$ is a local maximum, the Hessian matrix has exactly one negative eigenvalue, so the index $k=1$. In a neighborhood of $r_1$ this gives
\begin{equation}
T = T_1 - y^2.
\end{equation}
For $T<T_1$, in a neighborhood of $r_1$, the two distinct coordinate values $-y$ and $+y$ correspond to the same temperature, showing that multivaluedness already appears locally. Analogously, because $r_2$ is a local minimum, the index is $k=0$ and locally
\begin{equation}
T = T_2 + y^2.
\end{equation}
For $T>T_2$, a single temperature value again corresponds to two different coordinate values. These local expressions demonstrate that the inverse map $\pi^{-1}=r_+(T)$ is multivalued near each of the critical points $r_1$ and $r_2$, where the zero-rank of the differential $d\pi$ creates the branching of the inverse. At $T = T_1$, Branch1 and Branch2 merge, leading to a double root $r_1$. By continuity, Branch3 satisfies $r_3=\lim\limits_{T \to T_1}r_3(T) \neq r_1$ (if $r_3 = r_1$ held, this would contradict the assumption of two distinct critical points). Thus $T_1$ and $T_2$ serve as the branch points. Hence $r_+(T)$ is exactly two-valued on the closed interval $\mathcal{S}=[T_1,\,T_2]$ (one double root and one distinct root). A symmetric argument applies at $T_2$.

We now turn to establish the global multivaluedness of $r_+(T)$ from a geometric viewpoint using the theory of covering spaces. Since the map $\pi$ is not a local diffeomorphism at the critical points $r_1$ and $r_2$ of the manifold $\mathcal{M}$, we define $\widetilde{\mathcal{M}}$ as the topological space formed by the disjoint union of three open intervals
\begin{equation}
\widetilde{\mathcal{M}} = J_1 \sqcup J_2 \sqcup J_3,
\end{equation}
where
\begin{equation}
J_1 = (\mathcal{A},\,r_1),~J_2 = (r_1,\,r_2),~J_3 = (r_2,\,\mathcal{B}).
\end{equation}
The covering map $\widetilde{\pi} : \widetilde{\mathcal{M}} \rightarrow \mathbb{R}$ is defined by $\widetilde{\pi}(r_+)=T(r_+)$, where $\mathbb{R}$ is regarded as the base space. By the inverse function theorem, $\widetilde{\pi}$ is a local diffeomorphism in a neighborhood of any point $p$ on each branch. Applying the intermediate value theorem independently on each branch following the same steps as before, we can prove that the inverse map $\widetilde{\pi}^{-1}$ consists of three distinct values
\begin{equation}
\widetilde{\pi}^{-1}(T)= \{r_+^1,\,r_+^2,\,r_+^3\}.
\end{equation}

Now, consider a physical or geometric quantity $\mathcal{F} \colon \mathcal{M} \to \mathbb{R}$. Due to the multivaluedness of the inverse projection $\widetilde{\pi}^{-1}$, we can lift $\mathcal{F}$ to the covering space $\widetilde{\mathcal{M}}$ by defining the lifted function $\mathcal{F}^\uparrow \colon \widetilde{\mathcal{M}} \to \mathbb{R}$ as
\begin{equation}
\mathcal{F}^\uparrow \equiv \mathcal{F} \circ i_\mathcal{M},
\end{equation}
where $i_\mathcal{M} \colon \widetilde{\mathcal{M}} \to \mathcal{M}$ is the inclusion map. The lifted function $\mathcal{F}^\uparrow$ is single-valued on the covering space, on each branch of $\widetilde{\mathcal{M}}$ it gives a definite value. The quantity $\mathcal{F}$ regarded as a function of temperature is then defined by the composition
\begin{equation}
\mathcal{F}(T) = \mathcal{F}^\uparrow \circ \tilde{\pi}^{-1}(T).
\end{equation}
Because $\widetilde{\pi}^{-1}(T)$ is multivalued for $T\in (T_1,\,T_2)$, $\mathcal{F}(T)$ inherits this multivaluedness. Moreover, the physical or geometric quantities here are not constant functions, meaning that there exist branches $i\ne j$ and a temperature $T_0$ such that $\mathcal{F}^\uparrow[r_i(T_0)] \ne \mathcal{F}^\uparrow[r_j(T_0)]$. Hence these $\mathcal{F}(T)$ are naturally multivalued inside the spinodal region. Geometrically, when a phase transition occurs, the Morse lemma implies that the two nondegenerate critical points locally act as fold singularities of the temperature map. Consequently, the covering space $\widetilde{\mathcal{M}}$ is forced into a branched three-sheeted structure.

It is worth noting that at the critical point (where $\frac{\partial T}{\partial r_+} = \frac{\partial^2 T}{\partial r_+^2} = 0$), the two extrema merge into a single inflection point, and $r_+(T)$ becomes single‑valued. This corresponds to a cusp catastrophe, where the three branches merge into one.

\end{document}